\documentclass[structabstract]{aa}  
\usepackage{graphicx}
\usepackage{txfonts}
\begin{document}
   \title{WASP-4b Transit Observations With GROND\thanks{Based on observations
collected with the \textbf{G}amma \textbf{R}ay Burst \textbf{O}ptical 
  and \textbf{N}ear-Infrared \textbf{D}etector (GROND) at the MPG/ESO-2.2m telescope at La Silla Observatory, Chile. Programme 083.A-9010. }}

   \author{N. Nikolov \inst{1},
           Th. Henning\inst{1},
           J. Koppenhoefer\inst{2,3},
           M. Lendl\inst{4},
           G. Maciejewski\inst{5} \and 
           J. Greiner\inst{3}
}
           \institute{Max-Planck-Institut f\"{u}r Astronomie, K\"{o}nigstuhl 17, 69117 Heidelberg, Germany\\
           \email{nikolov@mpia.de}
           \and Universit\"{a}ts-Sternwarte M\"{u}nchen, Scheinerstr. 1, 81679 Munich, Germany
           \and Max Planck Institute for Extraterrestrial Physics, Geissenbachstr., 85748 Garching, Germany
           \and Observatoire de Gen\`{e}ve, Universit\`{e} de Gen\`{e}ve, 51 chemin des Maillettes, 1290 Sauverny, Switzerland
           \and Toru\'{n} Centre for Astronomy, Nicolaus Copernicus University, Gagarina 11, PL–87100 Toru\'{n} , Poland
}

\authorrunning{Nikolov et al.}
  \abstract
    {Ground-based simultaneous multiband transit observations allow an accurate system parameters determination and may lead to the detection and characterization of
     additional bodies via the transit timing variations (TTVs) method.}
    {We aimed to (i) characterize the heavily bloated WASP-4b hot Jupiter and its star by measuring system parameters and the 
    dependence of the planetary radius as a function of four (Sloan $g'$, $r'$, $i'$, $z'$) wavelengths and (ii) search for TTVs.}
    {We recorded 987 images during three complete transits with the GROND instrument, mounted on the MPG/ESO-2.2m telescope at La Silla Observatory. Assuming
a quadratic law for the stellar limb darkening we derive system parameters by fitting a composite transit light curve over all bandpasses simultaneously.
To compute uncertainties of the fitted parameters, we employ the Bootstrap Monte Carlo Method.}
    {The three central transit times are measured with precision down to 6 s. We find a planetary radius $R_{\mathrm{p}} = 1.413 \pm 0.020 R_{\mathrm{Jup}}$, 
an orbital inclination $i = 88.^{\circ}57 \pm 0.45^{\circ}$ and
calculate a new ephemeris, a period $P = 1.33823144 \pm 0.00000032$ days and a reference transit epoch $T_{\mathrm{0}} = 2454697.798311 \pm 0.000046$ (BJD). Analysis of the new 
transit mid-times in combination with previous measurements shows no sign of a TTV signal greater than 20 s. We perform simplified numerical simulations to place
upper-mass limits of a hypothetical perturber in the WASP-4b system.}
    {}



   \maketitle
%

\section{Introduction}
   More than 100 extrasolar planets have been detected to pass in front of the disk of its
   parent stars since the first transit observations reported nearly twelve years ago 
   (Charbonneau et al. 2000; Henry et al. 2000; Mazeh et al. 2000).
   These close-in planets orbit 
   their host stars with periods typically smaller than $\sim10$ days, probably formed at greater orbital distances
   and later migrated inward governed by processes which are still under debate. The discovery of
   each transiting extrasolar planet is of great interest for planetary science 
   as these objects provide a unique access to 
   an accurate determination of radii (down to a few percent) and masses via transit photometry and radial velocity measurements of the host star
   (Henry et al. 2000; Charbonneau et al. 2000). Moreover they allow one to plot thermal maps of the planetary surface via infrared spectra 
   (Richardson et al. 2007; Grillmair et al. 2007, Knutson et al 2007),
   determine the planetary temperature profiles and permit studies of the stellar spin-orbit alignment (Queloz et al. 2000, Triaud et al. 2010, Winn et al. 2011, Johnson et a. 2011). 
   All of these form a set of astrophysically precious parameters, which are
   critical for constraining the formation and evolution of these interesting objects.   
   Furthermore, the time intervals between successive transits are strictly constant if the system consists of
   a planet moving on a circular orbit around the parent star. If a third body is
   present in the system, it would perturb the transiting planet causing the time interval between successive transits to
   vary. The resulting transit timing variations (TTVs) allow the determination of the orbital period and mass of the perturber down to sub-Earth 
   masses (Miralda-Escud\'e 2002, Holman \& Murray 2005, Holman et al. 2010, Lissauer et al. 2011).  

   The transiting planet WASP-4b was discovered by Wilson et al. (2008) within the
   \textbf{W}ide \textbf{A}ngle \textbf{S}earch for \textbf{P}lanets in the southern hemisphere (WASP-S, Pollacco et al. 2006).
   The planet is a 1.12 $\rm{M_{Jup}}$ hot Jupiter orbiting a G7V star with a period of 1.34 day. WASP-4b
   was found to have a heavily irradiated atmosphere and inflated radius.
   Refined planetary orbital
   and physical parameters, based on transit photometry were presented by Gillon et al. (2009), who used the
   \bf{V}\rm{ery} \bf{L}\rm{arge} \bf{T}\rm{elescope} (VLT) to observe one transit, Winn et al. (2009), who measured
   two transits with the
   Magellan (Baade) 6.5m telescope at Las Campanas Observatory, Southworth et al.
   (2009), who observed four transits using the 1.54m Danish Telescope at ESO
   La Silla Observatory and
   Sanchis-Ojeda et al. (2011), who observed four transits using the Magellan
   (Baade) 6.5m telescope. During two of the transits,
   Sanchis-Ojeda et al. (2011) observed a
   short-lived, low-amplitude anomaly that the authors interpreted
   as the occultation of a starspot by the planet. Southworth et al. (2009) also noted similar anomalies in their light
   curves and the possibility that they were caused by starspot
   occultations. Sanchis-Ojeda et al. (2011), combined their data set
   with that of Southworth et al. (2009) and found that each of them is
   consistent with a single spot and a star that is well-aligned with the
   orbit. Tracking this starspot, it was possible to measure the rotation
   period of the host star. A new possible starspot has been recently
   detected in two closely spaced transits of WASP-4 by the MiNDSTEp
   collaboration with the Danish Telescope (Mancini 2011, private
   communication).
   Furthermore, Beerer et al. (2011) performed space-based secondary eclipse photometry,
   using the IRAC instrument on the Warm Spitzer Space Telescope, of the planet
   WASP-4b in the 3.6 and 4.5 $\mu$m bands. Their data suggest that
   WASP-4b's atmosphere lacks a strong thermal inversion on the
   day-side of the planet, which is an unexpected result for an highly irradiated
   atmosphere. C\'{a}ceres et al. (2011) analyzed high-cadence near-infrared ground-based
   photometry, detected the planet's thermal emission at 2.2 $\mu$m and concluded that WASP-4b shows 
   inhomogeneous redistribution of heat from its day- to night-side. Finally, Triaud et al. (2010) investigated
   the spin-orbit alignment ($\beta$) of the WASP-4 system by measuring the parent star radial velocity during
   a transit of its planet (Rossiter-McLaughlin effect) and found $\beta={4^{\circ}}^{+43^{\circ}}_{-34^{\circ}}$, excluding a projected retrograde orbit. 

   In this paper we present three new transits of WASP-4b observed in August and October 2009
   with the GROND instrument (Greiner et al. 2008), attached to the MPG/ESO 2.2m telescope at 
   La Silla Observatory. We recorded each transit in four optical $g'$, $r'$, $i'$ and $z'$ (Sloan)
   channels simultaneously. Using these new data sets we measure the planet orbital and physical parameters and
   constrain its ephemeris by fitting a new photometric data set over four pass-bands simultaneously. Unlike all previous
   observational studies related to the WASP-4b exoplanet, that fit data in a single bandpass, we model a transit
   light curve using the four bandpasses simultaneously. As pointed out by Jha et al. (2000), this approach permits one to break a fundamental
   degeneracy in the shape of the transit light curve. Each transit light curve can be described primarily by its depth and duration. For a single band
   observations it is always possible to fit these with a larger planet if the stellar radius is also increased and if the orbital inclination is decreased.
   The main advantage of the multi-band transit photometry is that it allows the determination of the planet orbital inclination unique, independent
   of any assumptions about the stellar and planetary radii. Finally, we add the three new mid-transit times and search for TTVs. This
   paper is organized as follows. In section 2 we present an overview of the three transit observations of WASP-4b and
   the data reduction. In section 3 we discuss the light curve analysis and the error estimation. Section 4
   summarizes the main results.

\section{Observations and data reduction}
  Three transits of WASP-4b were observed with the MPG/ESO 2.2-m telescope 
  at La Silla Observatory (Chile) during three runs on UT August 26 and 30 and
  October 8 2009 (see Fig. 1). Bad weather conditions prevented data collection during 
  the remaining two nights allocated for the project. To monitor the flux 
  of WASP-4, we used the \textbf{G}amma \textbf{R}ay Burst \textbf{O}ptical 
  and \textbf{N}ear-Infrared \textbf{D}etector (GROND). 
  It is a gamma-ray burst follow-up instrument, which allows 
  simultaneous photometric observations in four optical (Sloan $g'$, $r'$, $i'$, $z'$) 
  and three near-infrared ($JHK$) bandpasses (Greiner et al. 2008). In the optical channels
  the instrument is equipped with backside illuminated E2V CCDs ($2048\times2048$, 13.5 $\mu$m).
  The field of view for each of the four optical channels is $5.4'\times5.4'$ 
  with a pixel scale of $0''.158$ pixel$^{-1}$.

 \begin{figure*}[t]
   \centering
   \includegraphics[angle=0,width=16.5cm]{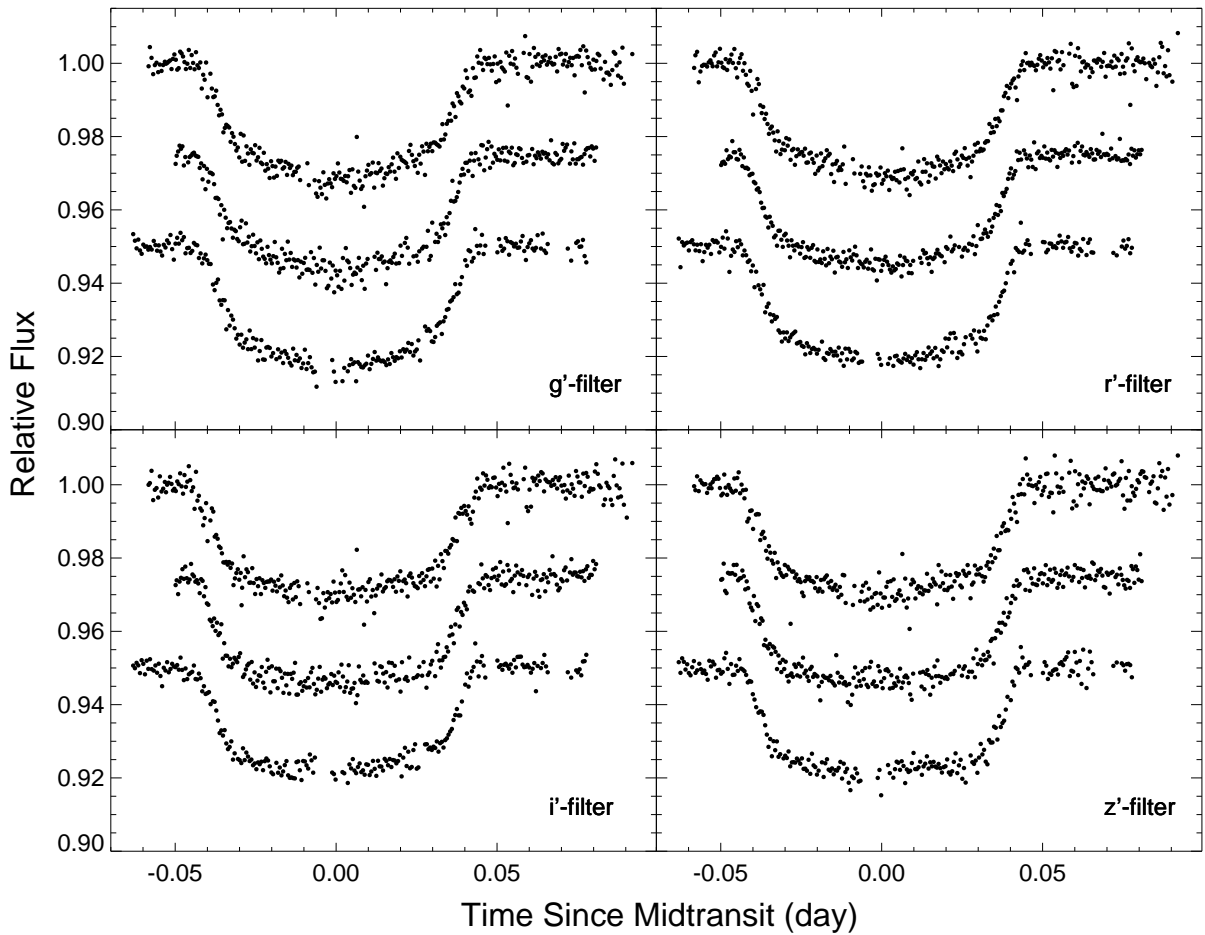}
      \caption{Relative $g'$, $r'$, $i'$ and $z'$ -band photometry of WASP-4 obtained with GROND. The transits
               occured (for each panel from top to bottom) on UT 2009 August 26, August 30 and October 8. The transit light curves
               obtained on each successive run are displayed with an offset of 0.025 in relative flux for a better illustration.}
         \label{FigVibStab5}
   \end{figure*}

  During each observing run
  we verified that WASP-4 and a nearby comparison star were
  located within GROND's field of view. The comparison star 
  was $\sim1.3'$ south-east from our target.
  For each run we obtained repeated integrations of WASP-4 and 
  the comparison star for $\sim 3.5$ hr, bracketing the predicted 
  central transit times. The exposure times were in the range 8--14 s, 
  depending on the weather conditions (i.e. seeing and transparency). During each run we used
  the fast readout mode ($\sim10$ s), achieving a cadence ranging from 18 to 24 s.  
  To minimize systematic effects in the photometry 
  associated with pixel-to-pixel gain variations we kept the telescope pointing stable. 
  However, due to technical issues caused by the guiding camera $\sim 35\%$ of the total number
  of images were acquired without guiding, that resulted in $\sim$ 10-12 pixel drifts per observing run ($\sim$ 3.5 hr).


  In the Near-Infrared (NIR) channels we obtained images with integration time of 10 s. 
  Longer exposures are not allowed due to technical constrains in the GROND instrument. It was found during the 
  analysis that the JHK-images were of insufficient signal-to-noise ratio (SNR) to be able to 
  detect the transit and to allow accurate system parameter derivation from the resulting light curves. In addition, the pixel 
  scale in the NIR detectors is 0.6 $\rm{px}^{-1}$, which resulted in a poorly sampled star 
  psfs (typically 4-5 pixels). As we carried out all observations in the NIR without 
  dithering to increase the cadence and to improve the time sampling and the photometry 
  precision, it is hardly possible to stack groups of NIR images with prior background 
  subtraction. Therefore, we excluded the NIR data in the further analysis.

  At the beginning of the observations on UT 2009 August 26 and 30,
  WASP-4 was setting from an air-mass of 1.02 and 1.05, respectively which
  increased monotonically to 1.5 and 1.6 at the end of the runs.
  During the October 8 observations, the air-mass decreased from 1.06 to 1.02 and
  then increased to 1.15 at the end of the run. 

  We employed standard IRAF\footnote{IRAF is distributed by the National Optical Astronomy
  Observatories, which are operated by the Association of Universities
  for Research in Astronomy, Inc., under cooperative agreement with the
  National Science Foundation.} procedures to perform bias and
  dark current subtraction as well as flat fielding. A median combined 
  bias was computed using 22 zero-second exposure frames and 4 dark frames were used 
  to compute the master dark. As non negligible fraction of the data was obtained 
  without guiding, it was critical to perform the data reduction with flat fields of high quality. A master 
  flatfield was calculated using the following methodology. From a set of 12 dithered twilight flats we selected frames with median pixel 
  counts in the range 10K to 35K which is well-inside the linear regime of the 
  GROND optical detectors. Finally, the reduced (bias- and dark-corrected) flat frames 
  were median combined to produce master flats for each of the g', r', i' and z' bands.

  Aperture photometry of WASP-4 and the reference star was 
  performed on each calibrated image. To produce the differential light curve
  of WASP-4, its magnitude was subtracted from that of the comparison star.
  Fortunately, the later was of similar color and brightness, which reduced the effect
  of differential color extinction. For example, the instrumental magnitude differences
  (comparison star minus WASP-4) measured on UT October 8 2009 were $\Delta g'=0.587$mag, $\Delta r'=0.669$mag, 
  $\Delta i'=0.681$mag, $\Delta z'=0.687$mag. As pointed out by Winn et al. (2009) the comparison star is slightly bluer than WASP-4 ($\Delta(g'-i')=-0.094$ mag).

  To improve the light curves quality we experimented with various aperture sizes and sky areas, aiming to
  minimize the scatter of the out-of-transit (OOT) portions, measured by the magnitude root-mean-square (r.m.s.).
  Best results were obtained with aperture radii of 16, 20.5 and 17.5 pixels for UT 2009 August 26 and 30 and October 8, respectively.

  The light curves contain smooth trends, most likely due to differential extinction. To improve
  its quality we decrease this systematic effect using the OOT portions of our light curves.
  We plot the magnitude vs. air-mass data for each run and channel and fit a linear model of the form:

  \begin{equation}
  f(z) = a + bz, 
  \end{equation}

  where $(a)$ and $(b)$ are constants and ($z$) is the air-mass. Once the coefficients were derived, we applied the correction to both the transit and
  the OOT data. 
  Performing this correction we notice that the baseline magnitudes of our target and the comparison star
  show slight, nearly linear correlations with the $(x, y)$ center positions of the PSF centroids on each CCD chip. 
  We remove these near-linear trends from WASP-4 and
  the reference star by modelling the base-line of the light curve with a polynomial that includes 
  a linear dependency on the centroid center positions $(x$ and $y)$ of the stars on the detectors using the function of the form:

  \begin{equation}
   f(x,y) = 1 + k_{x}x + k_{y}y + k_{xy}xy,
  \end{equation}

  where $(k_{x})$, $(k_{y})$ and $(k_{xy})$ are constants. We then
  subtracted the fit from the total transit photometry obtained for each channel during the three runs.
  No other significant trends that were correlated with instrumental parameters were found. 

   \begin{figure*}[t]
   \centering
   \includegraphics[angle=0,width=16.5cm]{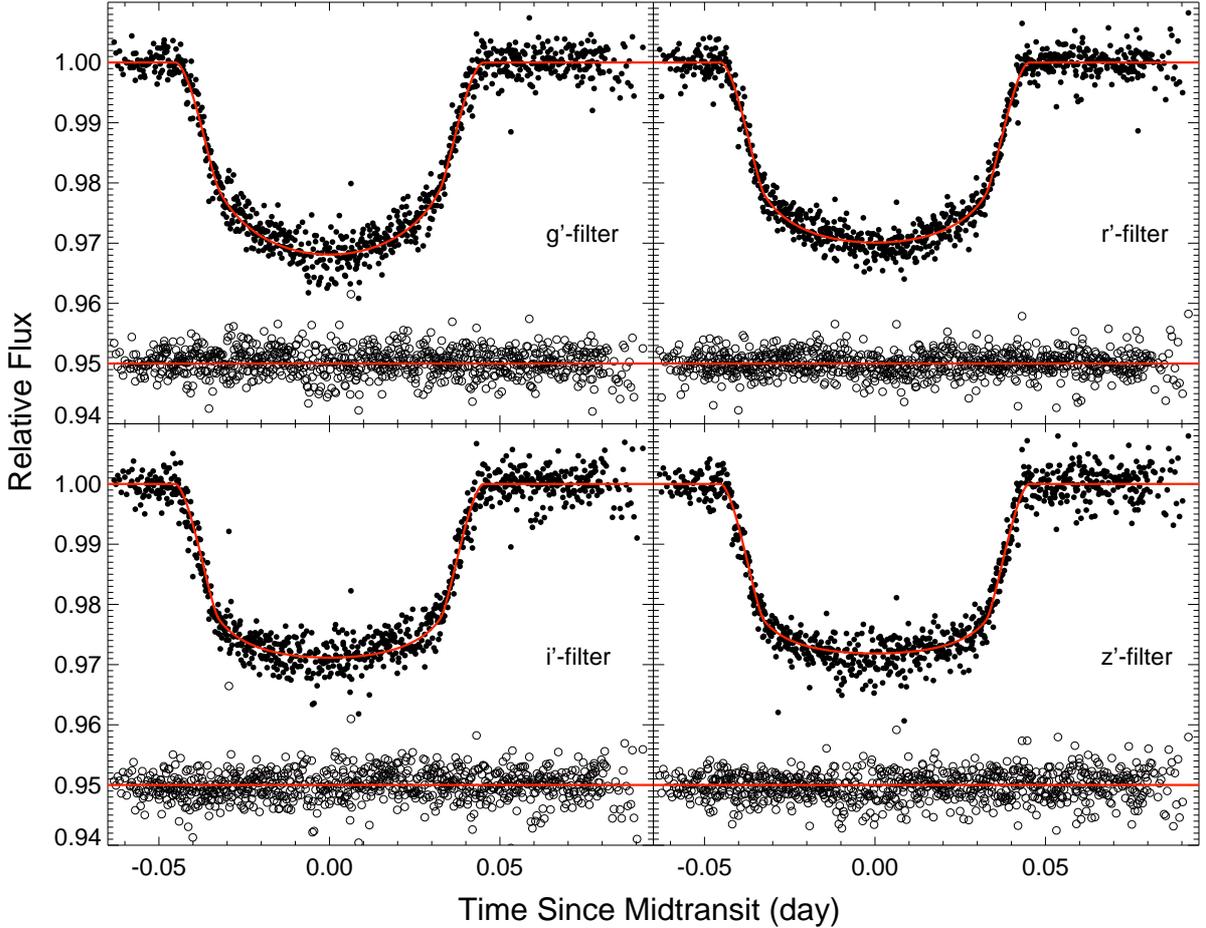}
      \caption{The composite light curves of WASP-4 obtained during three runs on UT 2009 August 26, August 30 and October 8. The best-fit transit 
               models are superimposed with continuous lines and the observed minus modeled residuals are shown centered at flux level 0.95 on each panel
               along a constant line. 
              }
         \label{FigVibStab5}
   \end{figure*}

  To illustrate the light curve quality improvement we present the scatter decrease after the described detrending procedure (Table 1).

\begin{table}
\caption{Light curve scatter difference measured with the standard deviation of the time series prior and after detrending. All quantities in mmag.}             
\label{table:1}      
\centering                          
\begin{tabular}{l c c c c}        
\hline\hline                 
run & $g'$ & $r'$ & $i'$ & $z'$\\    
\hline                        
  Aug 26      & 0.503  &  0.358 & 0.482  & 0.339  \\      
  Aug 30      & 0.818 & 0.973 & 0.633  & 0.725 \\      
  Oct 10      & 0.733  & 0.614 &  0.430  & 0.495 \\      
\hline                                   
\end{tabular}
\end{table}

  \section{Light curve analysis}
   To compute the relative flux of WASP-4 during transit as a function of the
   projected separation of the planet, we employed the models of Mandel \& Agol (2002), which
   in addition to the orbital period $P$ and the transit central time $T_C$ are a function of five parameters, including the planet to star 
   size ratio ($R_{\mathrm{p}}/R_{\mathrm{*}}$, where $R_{\mathrm{p}}$ and $R_{\mathrm{*}}$ are the absolute values 
   of the planet and star radii), the orbital inclination $i$,
   the normalized semimajor-axis ($a/R_{\mathrm{*}}$, where $a$ is the absolute value of the planet semimajor axis) and two limb darkening coefficients ($u_{1}$ and $u_{2}$).
   Because the parameters $a/R_{\mathrm{*}}$, $R_{\mathrm{p}}/R_{\mathrm{*}}$ and $i$ are degenerate in the transit light curve, we assume fixed values for 
   $M_{\mathrm{*}} = 0.92 \pm 0.06M_{\odot}$ from Winn et al. (2009) 
   and $R_{*}=0.907^{-0.013}_{+0.014} R_{*}$ from Sanchis-Ojeda et al. (2011) to break this degeneracy.
   Ideally we would like to fit for all parameters however, only the first three determine the best-fit radius of the star,
   the radius of the planet, its semimajor axis and inclination relative to the observer. 

   The values for the two limb darkening coefficients
   were included in our fitting algorithm to compute the theoretical transit models of Mandel \& Agol (2002). We assume the stellar limb darkening law to be quadratic,

  \begin{equation}
   \frac{I_{\mu}}{I_{1}} = 1 - u_{1}(1-\mu) - u_{2}(1 - \mu)^{2},
  \end{equation}

   where $I$ is the intensity and $\mu$ is the cosine of the angle between the line of sight and the normal to the stellar surface.
   We fixed (initially) the limb darkening coefficients ($u_{1}$ and $u_{2}$) to their theoretical values (see Table 2), which we obtained from the calculated and tabulated 
   ATLAS models (Claret 2004). Specifically, we performed a linear interpolation for WASP-4 stellar parameters $T_{\mathrm{eff}} = 5500\pm100$ K, $\log \, g  = 4.4813\pm0.0080$ 
   cgs, $[\mathrm{Fe/H}] = -0.03\pm0.09$ and 
   $v_{\mathrm{t}} = 2.0\pm1.0$ km $\rm s^{-1}$, which we adopted from Winn et al. (2009).

\begin{table}
\caption{Theoretical Limb-Darkening Coefficients (Quadratic Law).}             
\label{table:1}      
\centering                          
\begin{tabular}{l c c c c}        
\hline\hline                 
LD coefficient & $g'$ & $r'$ & $i'$ & $z'$\\    
wavelength (nm) & 455  & 627  & 763  & 893 \\    
\hline                        
   $u_{1}$ (linear)    & 0.623 & 0.413 & 0.314 & 0.248 \\      
   $u_{2}$ (quadratic) & 0.183 & 0.290 & 0.303 & 0.308  \\
\hline                                   
\end{tabular}
\end{table}

   To derive the best fit parameters we constructed a fitting statistic of the form:

  \begin{equation}
   \chi^{2} = \sum_{i=1}^{N_{f}} \left[ \frac{f_{i}(\rm{observed})-\it{f}_{i}(\rm{predicted})}{\sigma_{i}}\right] ^{2},
  \end{equation}

   where $f_{i}(\rm{observed)}$ is the flux of the star observed at the $i$-th moment (with the median of the OOT point normalized to unity), ${\sigma_{i}}$ controls
   the weights of the data points and $f_{i}(\rm{computed})$ is the predicted value for the flux from the theoretical transit light curve. We assume the orbital eccentricity
   to be zero and minimize the $\chi^{2}$ statistic using the downhill simplex routine, as implemented in the IDL
   AMOEBA function (Press et al. 1992). The minimization method evaluates iteratively the $\chi^{2}$ statistic, yet avoiding derivative estimation
   until the function converges to the global minimum.  

  In the entire analysis we derive uncertainties for the fitted parameters using the bootstrap Monte Carlo method (Press et al. 1992).
  It uses the original data sets (from each run and pass-band), with their $N$ data points to generate synthetic data sets also with the same 
  number of points with replacements. We then fit each data set to derive parameters. The process is repeated until we get an approximately 
  Gaussian distribution for each parameter and take the standard deviation of each distribution as the error of the corresponding fitted parameter.

\begin{table*}
\caption{System parameters of the WASP-4 system derived from the light curve analysis.}             
\label{table:3}      
\centering                          
\begin{tabular}{ l c c c c c}        
\hline\hline                 
Parameter & & Value & 68.3 $\%$ Conf. Limits  & Unit & Comment\\    
\hline                        
\hspace{0.3cm}\it{-- Light curve fit --}\\
Planet-to-star size ratio               & $p=R_{\mathrm{p}}/R_{\mathrm{*}}$         &  0.15655 &  0.00028 & --                 & 1\\
Orbital inclination                     & $i$                     &  88.57   &  0.45    & deg                & 1\\
Normalized semimajor axis               & $a/R_{\mathrm{*}}$               &  5.455   &  0.031   & --                 & 1\\
Transit impact parameter                & $b=(a/R_{\mathrm{*}})\cos{i}$    &  0.136   &  0.043   & --                 & 1\\
Reduced chi-square                      & $\chi^{2}_{\mathrm{red}}$&  1.005   &  --      & --                 & 1\\
\\
\hspace{0.3cm}\it{-- Planetary parameters --}\\
Radius                                  & $R_{\mathrm{p}} $                &  1.413   &  0.020   & $R_{Jup}$          & 2\\
Semimajor axis                          & $a$                     &  0.02300 &  0.00036 & AU                 & 2\\
Mean density                            & $\rho_{\mathrm{p}}$              &  0.582   &  0.036   & $\rho_{Jup}$       & 2\\
Surface gravity                         & $g_{\mathrm{p}}$                 &  15.36   &  0.91    & m $\rm{s}^{-2}$    & 2\\
Transit duration                        &  $t_{\mathrm{D}}$                &  2.1696  &  0.0047  & hour               & 1\\
Ingress/egress                          &  $t_{\mathrm{ing/egr}}$          &  0.3014  &  0.0039  & hour               & 1\\
\\
\hspace{0.3cm}\it{-- Stellar parameters --}\\
Mean density                            & $\rho_{*}$              &  1.715   &  0.029   & g $\rm{cm}^{-3}$   & 1\\
$g'$ Linear limb-darkening coefficient & $u_{\mathrm{1}}$                 &  0.616   &  0.046   & --                 & 1\\
$g'$ Quadratic limb-darkening coefficient & $u_{\mathrm{2}}$              &  0.211   &  0.060   & --                 & 1\\
$r'$ Linear limb-darkening coefficient     & $u_{\mathrm{1}}$             &  0.427   &  0.044   & --                 & 1\\
$r'$ Quadratic limb-darkening coefficient  & $u_{\mathrm{2}}$             &  0.303   &  0.064   & --                 & 1\\
$i'$ Linear limb-darkening coefficient & $u_{\mathrm{1}}$                 &  0.292   &  0.047   & --                 & 1\\
$i'$ Quadratic limb-darkening coefficient & $u_{\mathrm{2}}$              &  0.304   &  0.060   & --                 & 1\\
$z'$ Linear limb-darkening coefficient     & $u_{\mathrm{1}}$             &  0.241   &  0.045   & --                 & 1\\
$z'$ Quadratic limb-darkening coefficient  & $u_{\mathrm{2}}$             &  0.386   &  0.063   & --                 & 1\\
\hline                                   
\end{tabular}
\tablefoot{ (1) Based on the analysis of the multi-band light curves presented in this work; (2) Values obtained using (1) and results for the 
$M_{\mathrm{*}}$, $R_{\mathrm{*}}$, $T_{\mathrm{eff}}$ and $M_{\mathrm{p}}$
from Winn et al. (2009) and Sanchis-Ojeda et al. (2011).}
\end{table*}

   Ground-based time series data are often proned with time-correlated noise 
   (e.g. ``red noise''; see Pont et al. 2006) and therefore the data weights $\sigma_{i}$ need
   a special treatment, i.e. the uncertainties must be calculated accurately in order to obtain reliable estimates of the fitted parameters. 
   It is a common practice to use the calculated Poisson noise, or the observed standard deviation of
   the out-of-transit data for the weights. Our experience 
   shows that these methods often result in underestimated uncertainties of the
   modeled parameters. To estimate realistic parameter uncertainties
   we employ two methods. First we rescaled the photometric weights $\sigma_{j}$ so that the best-fitting model for each band and run results in 
   a reduced $\chi^2=1$, which therefore requires the initial values for the photometric uncertainties to be multiplied by the factors 
   $\chi^{2}_{\mathrm{red}} = 1.396$, $\chi^{2}_{\mathrm{red}} = 1.058$ and $\chi^{2}_{\mathrm{red}} = 1.179$, for
   the UT August 26 and 30 and October 8 2009 runs, respectively. 

   Second, we take into account the ``red noise" in our data by following the ``time-averaging" approach 
   that was proposed by Pont et al. (2006) and used in the transit data analysis of various authors including 
   Gillon et al. (2006), Winn et al. (2007, 2008, 2009) and Gibson et al. (2008). The main idea of the method
   is to compute the standard deviation (scatter) of the unbinned residuals between the observed and calculated fluxes, $\sigma_{1}$
   and also the standard deviation of the time-averaged residuals, $\sigma_{N}$, where the flux of each $N$ data points where averaged
   creating $M$ bins. In the absence of red noise one would expect
   
   \begin{equation}
         \sigma_{N}=\frac{\sigma_{1}}{\sqrt{N}}\sqrt{\frac{M}{M-1}}.
   \end{equation}

   However, in reality $\sigma_{N}$ is larger than $\sigma_{1}$ by some factor $\beta$, which, as pointed out by Winn et al. (2007), 
   is specific to each parameter of the fitted model. For simplicity we assume that the values of $\beta$ are the same for all parameters
   and find its value by averaging $\beta$ over a range of bins with timescales consistent with the duration of the transit ingress 
   or egress .i e. $10-30$ min. The resulting values for $\beta$ are then used to rescale $\sigma_{j}$ in the $\chi^{2}$ by this value (see Table 4) 
 
   To derive transit mid-times ($T_\mathrm{C}$) we fixed the orbital period $P$ to an 
   initial value taken from Sanchis-Ojeda et al. (2011) and fitted the light curves from each run for the 
   best $T_\mathrm{C}$, planet to star size ratio, inclination and normalized semimajor axis.
   We then find the best orbital period $P$ and $T_\mathrm{C}$ using the method discussed in 4.2. Then we
   take the new values for the $T_\mathrm{C}$ and $P$ and repeat the fit for the best planet to star size ratio,
   inclination and normalized semimajor axis. This procedure is iterated until we obtain a consistent solution.

\begin{table}
\caption{Summary of the residual r.m.s. (observed minus calculated flux) of the unbinned and binned data.}             
\label{table:2}      
\centering                          
\begin{tabular}{l c c c c}        
\hline\hline                 
run & band & $\sigma$ & $\beta$ & $\sigma_{sys}$\\    
    &      &   [mag]      &         & [mag]\\    
\hline                        
Aug 26 & $g'$ &   0.0025 &    1.41 &   0.0035\\
Aug 26 & $r'$ &   0.0025 &    1.38 &   0.0035\\
Aug 26 & $i'$ &   0.0028 &    1.45 &   0.0041\\
Aug 26 & $z'$ &   0.0028 &    1.49 &   0.0042\\
Aug 30 & $g'$ &   0.0023 &    1.19 &   0.0027\\
Aug 30 & $r'$ &   0.0017 &    1.10 &   0.0019\\
Aug 30 & $i'$ &   0.0025 &    1.23 &   0.0031\\
Aug 30 & $z'$ &   0.0022 &    1.29 &   0.0028\\
Oct 08 & $g'$ &   0.0018 &    1.22 &   0.0022\\
Oct 08 & $r'$ &   0.0018 &    1.17 &   0.0021\\
Oct 08 & $i'$ &   0.0019 &    1.29 &   0.0025\\
Oct 08 & $z'$ &   0.0020 &    1.33 &   0.0027\\
\hline                                   
\end{tabular}
\tablefoot{$\sigma$ represents the residual scatter over the entire
observing time interval on each run; $\sigma_{sys}$ is the rescaled value for
the photometric weights, reflecting the presence of red noise. }
\end{table}

  \section{Results}
  \subsection{System parameters}
   We set the mass and radius of the star equal to $0.92 \pm 0.06$ $M_{\odot}$ 
   and $0.907^{-0.013}_{+0.014}$ $R_{*}$, respectively, and determine the best-fit radius for WASP-4b, $R_{\mathrm{p}}$ by minimizing the $\chi^2$ function over the four
   band-passes and the three runs simultaneously. Furthermore, we assume that the inclination and the semimajor axis of the planetary orbit
   should not depend on the observed pass-band. Therefore, we constrained these parameters
   to a single universal value. Following this procedure for the radius of the planet we found $1.413\pm0.020$ $R_{\mathrm{Jup}}$ and the best-fit value for the 
   orbital inclination and semimajor axis $i=88.57^{\circ}\pm 0.45^{\circ}$ and $a = 0.02300 \pm 0.00036$ AU, respectively. 
   For comparison of the key parameter, $R_{\mathrm{p}}$, Wilson et al. (2008) derived $1.416^{+0.068}_{-0.043}$ $R_{\mathrm{Jup}}$, Gillon et al. (2008) 
   derived $1.304^{+0.054}_{-0.042}$ $R_{\mathrm{Jup}}$, Winn et al. (2009)
   derived $1.365\pm{0.021}$ $R_{\mathrm{Jup}}$, Southworth et al. (2009) derived $1.371^{+0.032}_{-0.035}$ $R_{\mathrm{Jup}}$ and Sanchis-Ojeda et al. (2011)
   derived $1.363\pm{0.020}$ $R_{\mathrm{Jup}}$. Our value for $R_{\mathrm{p}}$ is slightly higher than the cited radius in the literature and is dominated by the uncertainty of the stellar radius $R_{*}$. The
   value for the orbital inclination, from our analysis is in a good agreement with earlier results. For example, Winn et al. (2009) found ${88.56^\circ}^{-0.46}_{+0.98}$, 
   Sanchis-Ojeda et al. (2011) found ${88.80^\circ}^{-0.43}_{+0.61}$, while Gillon et al. (2009) and Soutworth et al. (2009) found
   ${89.35^\circ}^{+0.64}_{-0.49}$ and $88.80^\circ-90.00^\circ$, respectively. The results for the fitted parameters of the transit light curve modelling allow us to derive a 
   set of physical properties for the WASP-4 system. 
   Table 3 exhibits a complete list of these quantities.

   To complete the light curve analysis we further fit the light curves allowing the linear and the quadratic limb darkening coefficients ($u_1$ and $u_2$)
   in each pass-band to be treated as free parameters. We aim to compare the shapes of the transit light curves and the fitted system parameters from this
   fit and the curves derived using theoretically predicted limb darkening coefficients. In order to minimize the $\chi^2$-statistic using all of the 11 parameters, 
   including 8 limb darkening coefficients we again employed the downhill simplex algorithm and the bootstrap method. The new parameter values we find
   $R_{\mathrm{p}} = 1.409 \pm 0.020 R_{\mathrm{Jup}}$, $i = 88.75^{\circ} \pm 0.40^{\circ}$, $a = 0.02298 \pm 0.00033$ AU result in a slightly smaller value 
   of the $\chi^{2}_{\mathrm{red}}$ function (1.001), as the fit is able to remove some trends in the data. However,
   we report the parameter values derived using the theoretically predicted limb darkening coefficient as final results because
   the new fit is also more sensitive to systematics in the light curve photometry.

   A set of multi-band transit light curves also allows one to search for a dependence of the planetary radius, $R_{\mathrm{p}}$ as a function of the wavelength.
   Variations of $R_{\mathrm{p}}$ in some of the pass-bands could be produced by absorption lines such as water vapor, methane, etc. in the planetary atmosphere. 
   As a check for this assumption we fitted the data originating from each pass-band individually. Instead of fixing $R_{\mathrm{p}}/R_{\mathrm{*}}, i$ and $a/R_{\mathrm{*}}$ 
   to a a single universal value we set the 
   planet to star radius ratio as a free parameter and keep
   the remaining quantities as free parameters. Table 5 displays the best-fit radii as a function of the wavelength with no indications of variations within the measured errorbars.    

\begin{table}
\caption{Best-fit radius in each band.}             
\label{table:4}      
\centering                          
\begin{tabular}{c c c }        
\hline\hline                 
band & $\lambda$ & Radius  \\    
    &(nm)     & $R_{Jup}$\\    
\hline                        
$g'$&455 & $1.409 \pm 0.021$   \\
$r'$&627 & $1.415 \pm 0.020$   \\
$i'$&763 & $1.407 \pm 0.021$   \\
$z'$&893 & $1.420 \pm 0.021$   \\
\hline                                   
\end{tabular}
\tablefoot{The best-fit radii were derived after fixing the orbital inclination and normalized semimajor axis
to their best-fit values from Table 3.}
\end{table}

  \subsection{Deriving the transit ephemeris}
   One of our primary goals is to measure an accurate value for the transit ephemeris ($T_{0}$ and $P$). We include
   all available light curves from the three runs and fit
   for the locations of minimum light using the best-fit planetary radius,
   inclination and semimajor axis from Sect. (4.1.). We minimize the $\chi^{2}$ as defined in Sect. (4) over
   the four-band data of each run by fitting simultaneously for the $T_{\mathrm{C}}$. After the best-fit
   transit mid-times are derived we add them to all reported transit mid-times available at the time of writing in the 
   literature\footnote{Southwort et al. (2009) measured mid-times during four transits of WASP-4b. We exclude these times in our analysis,
   as they were reported as unreliable due to technical issues associated with the computer clock at the time of observations (Southworth 2011, private communication)}
   (Table 5). We further fit for the orbital period $P$ and the reference transit epoch $T_{\mathrm{0}}$ by plotting the transit mid-times
   as a function of the observed epoch $(E)$ 

  \begin{equation}
  T_{\mathrm{C}}(E) = T_{\mathrm{0}} + E\times P.
  \end{equation}

   In this linear fit the constant coefficient is
   the best-fit reference transit epoch ($T_{\mathrm{0}}$) and the slope of the line is the planetary period, $P$. 
   We then use the new values for the period and the reference epoch to repeat our fitting procedure for the
   system parameters until we arrive at a point of convergence.
   We used the on-line converter developed by Eastman et al. (2010) and transformed the transit mid-times 
   from JD based on UTC to BJD based on the Barycentric Dynamical Time (TDB). We derive a 
   Period $ = 1.33823144 \pm0.00000032$ day and a reference transit time $T_{\mathrm{C}} =  2454697.798311 \pm0.000046$ BJD.
   The result for the best-fit period is consistent with the one from Sanchis-Ojeda et al. (2011), who found $P = 1.33823187 \pm 0.00000025$ day,
   and reported a smaller error as they measured the period using all available 
   transit times, including the four additional measurements reported by Southworth et al. (2009).


\begin{table}
\caption{Literature transit mid-times of WASP-4 and their residuals in addition to the ephemeris derived 
in this work.}             
\label{table:1}      
\centering                          
\begin{tabular}{r l c c}        
\hline\hline                 
Epoch & Transit midtime & $\mathrm{O-C}$ & Reference \\    
      & (BJD) & (day) &  \\    
\hline                        
       0 & $2453963.1094^{+0.0025}_{-0.0021}$      &   0.00011& 1\\
     300 & $2454364.5765^{+0.0021}_{-0.0033}$      &  -0.00223& 1\\
     303 & $2454368.59341^{+0.00025}_{-0.00027}$   &   0.00003& 1\\
     305 & $2454371.26813^{+0.00097}_{-0.00087}$   &  -0.00171& 1\\
     324 & $2454396.69623^{+0.00015}_{-0.00026}$   &  -0.00001& 1\\
     549 & $2454697.798228^{+0.000055}_{-0.000055}$&  -0.00008& 2\\
     587 & $2454748.651228^{+0.000072}_{-0.000072}$&   0.00012& 2\\
     809 & $2455045.738643^{+0.000054}_{-0.000054}$&   0.00016& 3\\
     812 & $2455049.753274^{+0.000066}_{-0.000066}$&   0.00010& 3\\
     815 & $2455053.767816^{+0.000053}_{-0.000053}$&  -0.00006& 3\\
     827 & $2455069.826637^{+0.000086}_{-0.000086}$&  -0.00001& 4\\
     830 & $2455073.841103^{+0.000070}_{-0.000070}$&  -0.00024& 4\\
     850 & $2455100.605928^{+0.000061}_{-0.000061}$&  -0.00005& 3\\
     859 & $2455112.650002^{+0.000074}_{-0.000074}$&  -0.00005& 4\\
\hline                                   
\end{tabular}
\tablefoot{(1) Gillon et al. (2009); (2) Winn et al. (2009); (3) Sanchis-Ojeda et al. (2011); (4) This work; The epoch of the first observed transit of WASP-4b was taken as equal to zero.}
\end{table}

  We further investigated the observed minus calculated $(O-C)$ residuals of our data and
  the reported transit mid-times at the time of writing for any departures from the predicted 
  values estimated using our ephemeris. As it was shown by Holman \& Murray (2005) and 
  Agol et al. (2005) a deviation in the $O-C$ 
  values can potentially reveal the presence of moons or additional 
  planets in the system. We list and plot the $O-C$ values from our analysis in 
  Table 5 and Figure 3, respectively. An analysis of the $O-C$ values shows no 
  sign of a TTV signal greater than 20 s, except the two big outliers at epochs 300 and 305. We
  put upper constraints on the mass of an aditional perturbing planet in the system as a
  function of its orbital parameters. We simplified the three-body problem by assumimg
  that the system is coplanar and initial orbits of both planets are circular. The
  orbital period of the perturber was varied in a range between 0.1 and 10 orbital
  periods of WASP-4b. We generated 1000 synthetic $O-C$ diagrams based on calculations
  done with the \textsc{Mercury} package (Chambers 1999). We applied the
  Bulirsch--Stoer algorithm to integrate the equations of motion. Calculations covered
  1150 days, i.e. 860 periods of the transiting planet, that are covered by
  observations. The results of simulations are presented in Fig. 4. Our analysis
  allows us to exclude additional Earth-mass planets close to low-order period
  commensurabilities with WASP-4b.

   \begin{figure}
   \centering
   \includegraphics[angle=0,width=9.5cm]{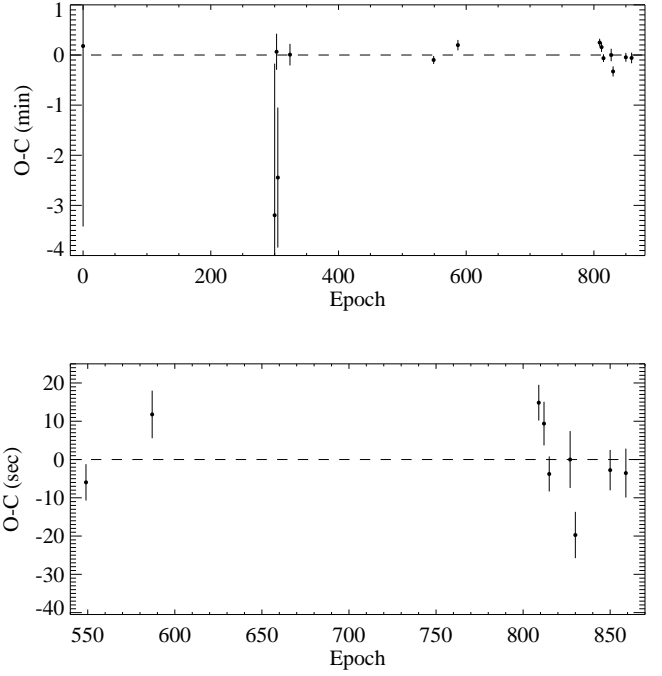}
      \caption{ The transit timing residuals for WASP-4b along with the one sigma errorbars. Top panel: The observed transit mid-times based on this work
                and others in the literature were subtracted from the calculated times produced by our ephemeris. Lower panel: A closer view
                of the available transit mid-times from 2009 and 2010. }
         \label{FigVibStab4}
   \end{figure}

   For the case of a transiting planet with a semi-major axis $a_{1}$ and period $P_{1}$ and a
  perturbing planet with semi-major axis $a_{2}$ on an outer orbit (i.e. $a_{2}\geq a_{1}$), period $P_{2}$ and mass $M_{2}$ Holman \& Murray 2005 derived the approximate
  formula

  \begin{equation}
   \Delta t \simeq \frac{45 \pi}{16} \left( \frac{M_{2}}{M_{*}} \right) P_{1} \alpha_{e}^{3} (1-\sqrt{2} \alpha_{e}^{3/2})^{-2}
  \end{equation}

  \begin{equation}
   \alpha_{e} = \frac{a_{1}}{a_{2}(1-e_{2})}
  \end{equation}

  for the magnitude of the variation (in seconds) of the time interval ($\Delta t$) between successive transits. One could imagine an exterior perturbing planet on a 
  circular coplanar orbit twice as far as WASP-4b (period $P\approx 3.75$ day not in mean motion resonance with the transiting planet). If such an imaginary 
  planet had a mass of $0.1 \rm{M_{Jup}}$, it would have
  induced $\sim 5$ sec variations in the predicted transit mid-times of WASP-4b. Furthermore,
  the perturber could cause radial velocity variations of the parent star $\sim13.78 $ m/s, which is
  below the radial velocity r.m.s. of the WASP-4 residual equal to 15.16 m/s, presented in the analysis of Triaud et al. (2010). Although,
  a few outliers are visible in the $O-C$ diagram we consider the prediction of such an imaginary planet in the WASP-4 system as
  immature, because the weights of the $O-C$ points might still require an additional term to account for systematics of unknown origin.

   \begin{figure}
   \centering
   \includegraphics[angle=0,width=8.0cm]{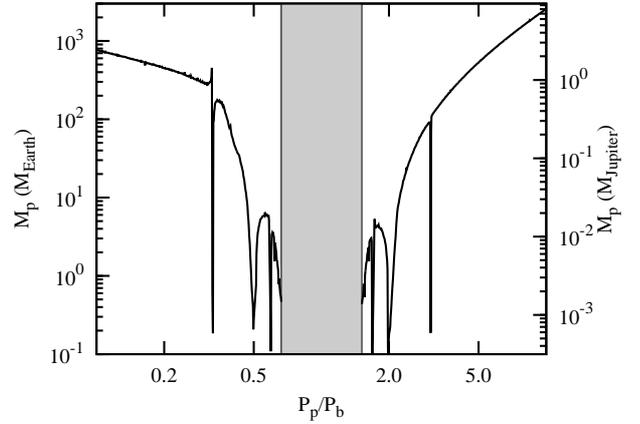}
      \caption{The upper-mass limit of a hypothetical additional planet that could perturb the
       orbital motion of WASP-4b as a function of ratio of orbital periods of transiting
       planet, $P_{\rm{b}}$, and the perturber, $P_{\rm{p}}$. Orbits located in a grey area
       were found to be unstable.}
         \label{FigVibStab4}
   \end{figure}

\section{Concluding remarks}
  We have used the GROND multi-channel instrument to obtain four-band simultaneous light curves of the WASP-4 system
  during three transits (a total of 12 light curves) with the aim to refine the planet and star parameters and to search
  for transit timing variations. We derived the final values for the planetary radius $R_{\mathrm{p}}$ and the orbital
  inclination $i$ by fixing the stellar radius $R_{*}$ and mass $M_{*}$ to the independently derived values of Winn et al. (2009) and Sanchis-Ojeda et al. (2011).
  We include the time-correlated ``red-noise" in the photometric uncertainties using the ``time-averaging" methodology and
  and by rescaling the weights to produce value for the $\chi^{2}_{\mathrm{red}}$ equal to unity. We further 
  perform the light curves analysis by minimizing the $\chi^{2}_{\mathrm{}}$ function over all pass bands and runs
  simultaneously via two approaches. First we modeled the data using theoretically predicted limb-darkening
  coefficients for the quadratic law. Second, we fit the light curves for the limb-darkening. Both methods result in
  consistent system parameters within $<1\%$. The second method produced limb darkening parameters compatible with the theoretical predictions within the one-sigma errorbars of the fitted parameters.

 We added three new transit mid-times for WASP-4b, derived a new ephemeris and investigated the $O-C$ diagram for outlier points. We have not found compelling
 evidences for outliers that could be produced by the presence of a second planet in the system. We did not detect any short-lived photometric
 anomalies such as occultations of starspots by the planet, which where detected by Sanchis-Ojeda et al. (2011). At the transit r.m.s. level of our
 light curves ($\sim2.2$ mmag) it would be challenging to detect similar anomalies.
 However we note that due to the brightness of the parent star, the short 
 planetary orbital period and the significant transit depth, the WASP-4 system is well suited for follow-up observations.

\begin{acknowledgements}Part of the funding for GROND (both hardware as well as personnel) was 
   generously granted from the Leibniz-Prize to Prof. G. Hasinger 
   (DFG grant HA 1850/28-1). N.N. acknowledges the Klaus Tschira Stiftung (KTS) and the Heidelberg 
   Graduate School of Fundamental Physics (HGSFP) for the financial support of his PhD research. 
   GM acknowledges the financial support from the Polish Ministry of Science and Higher
   Education through the Iuventus Plus grant IP2010 023070.  The authors would like
   to acknowledge Luigi Mancini, John Southworth and the anonymous referee by their useful comments and suggestions.
\end{acknowledgements}

\newpage

\end{document}